\documentclass[12pt]{article}
\usepackage{amsmath,amssymb,amsthm,amsfonts,url}

\newcommand{\be}{\begin{equation}}
\newcommand{\ee}{\end{equation}}
\newcommand{\ba}{\begin{eqnarray}}
\newcommand{\ea}{\end{eqnarray}}
\newcommand{\ban}{\begin{eqnarray*}}
\newcommand{\ean}{\end{eqnarray*}}

\newcommand{\sandwich}[3]{\mbox{$ \langle #1 | #2 | #3 \rangle $}}
\newcommand{\ket}[1]{\mbox{$ | #1 \rangle $}}
\newcommand{\bra}[1]{\mbox{$ \langle #1 | $}}

\newcommand{\si}{\sigma}
\newcommand{\demi}{\frac{1}{2}}

\newcommand{\real}{\begin{picture}(8,8)\put(0,0){R}\put(0,0){\line(0,1){7}}\end{picture}}

\newcommand{\one}{\leavevmode\hbox{\small1\normalsize\kern-.33em1}}

\newcommand{\CHSH}{\text{\textit{CHSH}}}

\begin{document}

\title{An anomaly of non-locality}

\author{Andr\'e Allan M\'ethot and Valerio Scarani\\[0.5cm]
\normalsize\sl Group of Applied Physics\\[-0.1cm]
\normalsize\sl University of Geneva, rue de l'Ecole-de-M\'edecine 20\\[-0.1cm]
\normalsize\sl 1211 Geneva 4~~\textsc{Switzerland}}

\maketitle

\begin{abstract}
Ever since the work of Bell, it has been known that entangled
quantum states can rise non-local correlations. However, for almost
forty years, it has been assumed that the most non-local states
would be the maximally entangled ones. Surprisingly it is not the
case: non-maximally entangled states are generally more non-local
than maximally entangled states for all the measures of non-locality
proposed to date: Bell inequalities, the Kullback-Leibler distance,
entanglement simulation with communication or with non-local boxes,
the detection loophole and efficiency of cryptography. In fact, one
can even find simple examples in low dimensions, confirming that it
is not an artefact of a specifically constructed Hilbert space or
topology. This anomaly shows that entanglement and non-locality are
not only different concepts, but also truly different resources. We
review the present knowledge on this anomaly, point out that Hardy's
theorem has the same feature, and discuss the perspectives opened by
these discoveries.
\end{abstract}

\section{Introduction}\label{sec:intro}

The history of quantum non-locality is far from being continuous: it
is indeed made of abrupt steps followed by periods of stagnation
(which are however becoming shorter and shorter in the recent few
years). The field was initiated in 1935 by Einstein, Podolsky and
Rosen (EPR) who used entanglement in a cleverly constructed argument
to attack the validity of quantum physics as a complete theory of
Nature~\cite{epr35}. An entangled state was specified such that,
when a position measurement is made on the first particle, the
position of the second is known with perfect predictability and,
conversely, when a momentum measurement is made on the first
particle, the momentum of the second particle is known with
arbitrary precision. Ruling out a \emph{spukhafte Fernwirkung}
(spooky action at a distance), EPR concluded that position and
momentum must be elements of reality, i.e. must have values
predetermined before the measurement. If translated into a
mathematical formalism, local realism---the point of view put forth
by EPR--- translates into local hidden variable (LHV) models. It
took the better part of three decades for Bell to come along and
realize that if LHVs are indeed present, then the predictions of
quantum theory cannot be correct~\cite{bell64}. Bell's result opens
the possibility of discriminating experimentally between LHVs or
quantum theory; still, not many people rushed on the test, and about
20 years had to elapse before the issue was settled (at least for
the majority of the physicists) in favor of quantum theory
\cite{aspect}. Experiments have multiplied since, but we won't focus
on them and come rather back to theory.

Letting apart the appearance of the Greenberger-Horne-Zeilinger
argument involving more than two particles~\cite{ghz89}, we can
safely say that until 1989 the studies on non-locality had focused
on a single quantum state, namely the singlet state of two spins
one-half (qubits), which is in modern notations
\begin{equation}
\ket{\Phi^+}=\frac{1}{\sqrt{2}}\,\left(\ket{00} +\ket{11}\right).
\label{eq:singlet}
\end{equation}
In 1991, Gisin asked whether any other bipartite entangled state
could be non-local, and he found that {\em all} pure entangled
bipartite states are in fact non-local \cite{gisin}. Shortly later,
Popescu and Rohrlich demonstrated that any pure entangled quantum
state is non-local \cite{pr92}. For mixed states, the situation is
more complex; two milestones are~\cite{werner,horo}, but we won't
discuss these issues here.

With the advent of quantum information, around 1995, the study of
entanglement accelerated rapidly. One of the new ideas was to define
a quantitative measure for entanglement. This problem is still
unsolved in general. However, for pure bipartite quantum states (on
which we focus from now on), the amount of entanglement is uniquely
defined as
\begin{equation}
{\cal{E}}\left(\ket{\psi_{AB}}\right)=S(\rho_A)\label{entg}
\end{equation}
where $S(\cdot)$ is von Neumann entropy and
$\rho_A=\text{Tr}_B(\ket{\psi_{AB}}\bra{\psi_{AB}})$ is the state of
$A$ obtained by partial trace~\cite{dhr02}. Note that since
$S(\rho_A)=S(\rho_B)$ for all bipartite pure states, the definition
is not ambiguous. As a consequence, a maximally entangled state of
two $d$-level systems is one which can be written (in a convenient
basis) as
\begin{equation}
\ket{\Psi_d}=\frac{1}{\sqrt{d}}\,\sum_{k=0}^{d-1} \ket{k}\ket{k},
\end{equation}
since it reaches up to
${\cal{E}}\left(\ket{\Psi_d}\right)=\log_2(d)$ bits. In particular,
the state $\ket{\Phi^+}$, see~\eqref{eq:singlet}, is a maximally
entangled state of two qubits.

In this passion for entanglement, non-locality was however left
aside from the mainstream. This was the reason why, apart from a
result by Eberhard in 1993 \cite{eberhard93} which went almost
unnoticed, it took another decade after Gisin's 1991 theorem to
unlock another feature of non-locality. At first sight, this feature
appears as an anomaly: {\em for all measures of non-locality
invented to date, it happens almost always that the most non-local
state is not the maximally entangled one}. It is the object of the
present paper to review the evidences collected to date for this
anomaly, to see what can be deduced from this anomaly and to point
the possible new directions of research that it inspires.

\section{The basic case shows no anomaly}\label{sec:basic}

A good starting point consists in reviewing first Gisin's theorem
\cite{gisin}, thus showing a case where the anomaly does \emph{not}
show up.

\subsection{The CHSH inequality}

Let's introduce first the basic tool: the Bell inequality derived by
Clauser, Horne, Shimony and Holt (CHSH) in 1969 \cite{chsh}. Let
$A_{1}$ and $A_{2}$ be two possible measurements on particle $A$,
whose outcomes are written $a_1$ and $a_2$ In this setting, the
outcomes are binary and we write them as $a_j\in\{-1,+1\}$. Let
similar definitions hold for the measurements $B_{1}$ and $B_{2}$ on
particle $B$. Let us now define, in this context, an element of
reality $\lambda$ has an ensemble ensemble of possible answers, the
quadruple $(a_1,a_2,b_1,b_2)$, to any measurements on particle $A$
and $B$. One can then compute the function $\CHSH_L(\lambda) =
a_1b_1 + a_1b_2 + a_2b_1 - a_2b_2$ and convince himself that $-2\leq
\CHSH_L(\lambda) \leq 2$ for any choice of $\lambda$. It is
important to note that the value of the function $\CHSH_L(\lambda)$
cannot be determined directly by experimentations, for $a_1$ and
$a_2$ cannot be measured simultaneously. The same holds for $b_1$
and $b_2$. However, by measuring several identically produced pairs
with randomly chosen measurements, one can estimate the average
value
\begin{equation}
\begin{aligned}\label{chshgen}
\CHSH_L\equiv & \int \rho(\lambda)\CHSH_L(\lambda)\,d\lambda\\
=& E(A_{1},B_{1}) + E(A_{1},B_{2}) + E(A_{2},B_{1}) -
E(A_{2},B_{2}),
\end{aligned}
\end{equation}
where $E(\cdot)$ is the expectation value and $\rho(\lambda)$ is the
probability distribution of $\lambda$. The CHSH inequality is just
the obvious conclusion
\begin{equation}
\left|\CHSH_L\right| \leq  2.
\end{equation}
The CHSH inequality is
remarkable since it is the unique Bell inequality when restricting
to two possible measurements per particle and binary
outcomes~\cite{fine,cg}. Let us now turn our attention to the
quantum world.

\subsection{Gisin's theorem}

The most general pure state of two qubits can be written in the
Schmidt basis as
\begin{equation}\label{eq:psitheta}
\ket{\psi(\theta)} = \cos\theta\ket{00}+ \sin\theta\ket{11}
\end{equation}
with $\theta\in[0,\frac{\pi}{4}]$. Through all this paper, when we
describe qubits, we use the convention that $\ket{0}$ and $\ket{1}$
are the eigenstates of the Pauli matrix $\si_z$ for the eigenvalue
$+1$ and $-1$ respectively\footnote{When measured in the calculation
basis, $\sigma_z$, the $\ket{0}$ and $\ket{1}$ state return the
values $+1$ and $-1$ respectively. The usual measurement rules for a
state in superposition of $\ket{0}$ and $\ket{1}$ applies.}. Any
projective measurement on a qubit can be described by the projection
on the eigenstates of a Pauli matrix $\vec{n}\cdot\vec{\sigma}$
where $\vec{n}$ is a normalized unit vector. Therefore we can
rewrite Equation~(\ref{chshgen}) as
\begin{equation}\label{chshq}
\CHSH_Q(\{\vec{a}_{i},\vec{b}_{j}\})\equiv E(\vec{a}_{1},\vec{b}_{1}) +
E(\vec{a}_{1},\vec{b}_{2}) + E(\vec{a}_{2},\vec{b}_{1}) -
E(\vec{a}_{2},\vec{b}_{2}),
\end{equation}
where a simple quantum mechanical calculation yields
\begin{equation}\label{eq:eab}
\begin{aligned}
E(\vec{a}_{i},\vec{b}_{j})= &\
\sandwich{\psi(\theta)}{(\vec{a}_i\cdot\vec{\sigma})
\otimes(\vec{b}_j\cdot\vec{\sigma})}{\psi(\theta)}\\
= &\ a_z^ib_z^j\,+\,\sin(2\theta)\, \big(a_x^ib_x^j -
a_y^ib_y^j\big).
\end{aligned}
\end{equation}
We can then put this expression into~\eqref{chshq} and try and
maximize the expression by choosing the measurements conveniently.
The maximum value is found when one chooses $\vec{a}_1=\hat{z}$,
$\vec{a}_2=\hat{x}$, $\vec{b}_1=\frac{1}{\sqrt{2}}(\hat{z}+\hat{x})$
and $\vec{b}_2=\frac{1}{\sqrt{2}}(\hat{z}-\hat{x})$. This value is
\begin{equation}\label{violtheta}
\CHSH_Q(\theta) = 2
\sqrt{1+\sin^2(2\theta)}
\end{equation}
which is always larger than
2 unless $\theta=0$. In conclusion, any pure state of two qubits
(but obviously the separable ones) violates the CHSH inequality.

\subsection{Comparison with entanglement}

The amount of entanglement contained in $\ket{\psi(\theta)}$ is
readily computed: from
\begin{equation}
\rho_A = \rho_B = \left(\begin{array}{cc}\cos^2\theta & 0\\
0 & \sin^2\theta\end{array}\right)
\end{equation}
one obtains
\begin{equation}
{\cal{E}}\left(\ket{\psi(\theta)}\right) = H\left(\cos^2\theta,
\sin^2\theta\right)
\end{equation}
where $H(\cdot)$ is the well known
Shannon entropy. In the region $\theta\in[0,\frac{\pi}{4}]$, the
entropy is increasing in function of $\theta$, as expected. Such is
also the violation $\CHSH_Q(\theta)$ given in \eqref{violtheta}.
Consequently, in the example studied here, the more entangled is a
state, the more non-local it is, which seems pretty natural.
However, we shall see that such a state of affairs is rather an
exception!

\section{Escaping the detection loophole}\label{sec:dlh}

We first review the issue of the detection loophole; being the very
first example in which the anomaly of non-locality showed
up~\cite{eberhard93}. We consider the same physical situation as in
the previous Section, namely two binary measurements on each qubit
of a pair. Ideally, for any measurement, only the two results $+$
and $-$ are possible. However, in a true experiment there is a third
possible result, labelled $\perp$, corresponding to the events where
the detector do not fire. Physicists normally make the natural
assumption of fair sampling: the particles that are detected
constitute a representative set of all the particles. In other
words, the fact that the detector does not fire depends on
parameters which are completely uncorrelated to the two-qubit state
that is measured. The assumption of fair sampling is reasonable and
in agreement with the orthodox understanding of quantum physics.
Nonetheless, when key issues like non-locality are at stake, it is
also reasonable to be careful.

To study the detection loophole, one must rewrite the CHSH
inequality in a form first derived by Clauser and Horne~\cite{ch}:
\begin{equation}\label{eq:ch}
\begin{aligned}
\Pr\nolimits_{A_1B_1}[++]+\Pr\nolimits_{A_1B_2}[++]+\Pr\nolimits_{A_2B_1}[++]-\Pr\nolimits_{A_2B_2}[++]& \\
-\Pr\nolimits_{A_1}[+]-\Pr\nolimits_{B_1}[+] & \leq 0.
\end{aligned}
\end{equation}
This inequality can be obtained directly from Equation~(13)
of~\cite{eberhard93} by noticing that
\begin{equation}
\begin{aligned}
\Pr\nolimits_{A_1B_2}[+-] + \Pr\nolimits_{A_1B_2}[+\negthinspace\negthinspace\perp]
 = &\Pr\nolimits_{A_1}[+] - \Pr\nolimits_{A_1B_2}[++]\\
\Pr\nolimits_{A_2B_1}[-+] +
\Pr\nolimits_{A_2B_1}[\perp\negthinspace\negthinspace+] =
&\Pr\nolimits_{B_1}[+] - \Pr\nolimits_{A_2B_1}[++].
\end{aligned}
\end{equation}
Quantum mechanics tells us that if we are to perform measurements on
the state $\ket{\psi(\theta)}$ with detectors of efficiency $\eta$,
we would have
\begin{equation}
\begin{aligned}
\Pr\nolimits_{A_i}[+]=&\ \eta\,\demi\left(1+\cos (2\theta)\,a_z^i\right),\\
\Pr\nolimits_{B_i}[+]=&\ \eta\,\demi\left(1+\cos (2\theta)\,b_z^i\right)\ \text{and}\\
\Pr\nolimits_{A_iB_j}[++]=&\ \eta^2\,\frac{1}{4}\left(1+\cos
(2\theta)\,(a_z^i+b_z^j)+E(\vec{a}_{i},\vec{b}_{j})\right),
\end{aligned}
\end{equation}
where $E(\vec{a}_{i},\vec{b}_{j})$ is defined in
Equation~\eqref{eq:eab}. By reinserting these expressions into the
inequality~\eqref{eq:ch}, one finds that the inequality can be
violated if and only if the efficiency of the detector is high
enough, namely
\begin{equation}
\eta >
\eta_c(\theta)\,=\,\min_{\{\vec{a}_{i},\vec{b}_{j}\}}\left[\frac{4+2\cos
(2\theta)\,(a_z^1+b_z^1)}{2+2\cos
(2\theta)\,(a_z^1+b_z^1)+\CHSH_Q(\{\vec{a}_{i},\vec{b}_{j}\})}\right].
\end{equation}
For the present study, we want to see for which state the closure of
the detection loophole requires the smallest detection efficiency.
This criterion can be seen as a measure of non-locality, since
(intuitively) the more non-local a state is, the easier its
non-locality is to be revealed in an imperfect measurement. If we
now recall the settings that maximized
$\CHSH_Q(\{\vec{a}_{i},\vec{b}_{j}\})$ from the previous Section, we
had $\CHSH_Q(\theta)=2\sqrt{1+\sin^2(2\theta)}$. It is then easy to
verify that, as $\theta$ increases, $\eta_c(\theta)$ decreases, with
the minimum obtained at $\theta = \pi/4$. It thus seems that
everything is as it should be. Nonetheless, it is not equivalent to
maximize the function $\CHSH_Q(\{\vec{a}_{i},\vec{b}_{j}\})$ and to
minimize $\eta_c(\theta)$ directly. For any $\theta<\frac{\pi}{4}$,
$\eta_c(\theta)$ is minimized by settings which are \emph{not} those
that maximize the violation of the CHSH inequality (even though they
still give a violation), and moreover, with the optimized settings,
$\eta_c(\theta)$ decreases as $\theta$ decreases. In particular, one
has $\eta_c(\theta\rightarrow 0)\rightarrow \frac{2}{3}$.

One might think that this astonishing unusual characteristic is not
an anomaly of non-locality or of entanglement, but of the specific
inequality that has been considered. However, there exists an
explicit local model which recovers the quantum predictions for the
maximally entangled state (independently of any inequality) as soon
as $\eta\leq\frac{3}{4}$~\cite{gisin2}. This means that for
$\frac{2}{3}<\eta\leq\frac{3}{4}$, the maximally entangled state can
in no way close the detection loophole, while some non-maximally
entangled states can.

\section{Two qutrits, two measurements}\label{sec:klcrypto}

In the case of the detection loophole, we have seen that the anomaly
arises for the simplest case of composed systems, namely two
qubits\footnote{Other appearances of the anomaly for two qubits will
be discussed in Section \ref{sec:sim} and~\ref{sec:hardy}.}. In this
Section, we discuss measures of non-locality for which the anomaly
is not present for two qubits, but it appears for the next simplest
case, namely when the non-locality of two {\em three-level systems
(qutrits)} is studied using only two settings per qutrit.

To understand this Section, a sketch of the geometrical view of the
problem may be useful. Once the number of parties (here, two), of
settings (here, two per party) and the dimension of the outcomes
(here, a trit per party) is fixed, one can represent all possible
probability distributions as a closed set in a large-dimensional
space; the borders are given by the constraints that all
probabilities must be positive and sum up to one. Within this set,
one can define the set of local correlations: this set is closed,
convex, and has a finite number of extremal points (that is, all
local distributions can be written as convex combinations of those
points). Technically, such a set is called a \emph{polytope}. A
point outside the polytope of local correlations represents, quite
obviously, a non-local probability distribution. In this view, a
Bell inequality is a facet of the polytope, and to violate the
inequality means precisely that the point lies above the facet, i.e.
outside the polytope. The amount of violation is related to the
geometrical distance between the facet and the point, and is thus a
natural candidate for a measure of non-locality\footnote{It is to be
noted that it is this exact measure, termed differently, that we
have used in Section~\ref{sec:basic}. It should also be noted that
this measure is not unique: as we will see, many measures of
non-locality have been put forth. They all seem \emph{a priori} as
good one another.}.

\subsection{Violation of Bell's inequality}\label{sec:bellineq}

We focus on Bell inequalities for two qutrits using two settings per
qutrit. In this case, as for two qubits, it has been proved
\cite{cg,masanes} that all the facets of the polytope are equivalent
up to trivial symmetries (like relabelling of the settings and of
the outcomes). In other words, a single Bell inequality is
meaningful, which is the so-called CGLMP inequality~\cite{cglmp}:
\begin{equation}\label{eq:i3}
\begin{aligned}
\text{\textit{CGLMP}}_L= &\ \Pr[a_1\equiv b_1]+\Pr[a_1\equiv b_2]+\Pr[a_2\equiv b_1]\\
&\ +\Pr[a_2\equiv b_2+2]-\Pr[a_1\equiv b_1+1]-\Pr[a_1\equiv b_2+2]\\
&\ -\Pr[a_2\equiv b_1+2]-\Pr[a_2\equiv b_2]\leq 2,
\end{aligned}
\end{equation}
where $a_i,b_j\in\{0,1,2\}$ and the equivalence is modulo 3. The
study of this inequality in general is tedious, all the same a lot
of symmetry is found if we restrict to the settings that a thorough
study demonstrated to be the optimal ones for the cases discussed
here\footnote{This search through all possible settings has always
been performed numerically~\cite{a,g}.}. We can write the optimal
settings explicitly in the form of projectors onto the following
states\footnote{These projectors are equivalent as first applying a
phase shift ($\ket{0}\rightarrow \ket{0}, \ket{1}\rightarrow
e^{i\alpha_j}\ket{1}$ and $\ket{2}\rightarrow
e^{2i\alpha_j}\ket{2}$; likewise for Bob with
$\alpha_j\rightarrow\beta_j$), then the Fourier transform and
finally measuring in the computational basis.}: \ba A_j&:&
\left\{\begin{array}{lcl}\ket{a_j=0}&=&\ket{0}+e^{i\alpha_j}\ket{1}+
e^{2i\alpha_j}\ket{2}\\
\ket{a_j=1}&=&\ket{0}+\chi e^{i\alpha_j}\ket{1}+
\bar{\chi}e^{2i\alpha_j}\ket{2}\\
\ket{a_j=2}&=&\ket{0}+\bar{\chi}e^{i\alpha_j}\ket{1}+ \chi
e^{2i\alpha_j}\ket{2}
\end{array}\right.\text{and} \\ B_k&:&
\left\{\begin{array}{lcl}
\ket{b_k=0}&=&\ket{0}+e^{i\beta_k}\ket{1}+
e^{2i\beta_k}\ket{2}\\
\ket{b_k=1}&=&\ket{0}+\bar{\chi} e^{i\beta_k}\ket{1}+
{\chi}e^{2i\beta_k}\ket{2}\\
\ket{b_k=2}&=&\ket{0}+{\chi}e^{i\beta_k}\ket{1}+ \bar{\chi}
e^{2i\beta_k}\ket{2}
\end{array}\right.,\ea
where we have omitted the normalization factors $1/\sqrt{3}$ for
readability, have defined $\chi=e^{2i\pi/3}$ and
$\bar{\chi}=e^{-2i\pi/3}$ and left $\alpha_j$ and $\beta_k$
undefined for now\footnote{Note that the role of $\chi$ and
$\bar{\chi}$ is reversed between A and B.}. Let us consider a state
which is Schmidt-diagonal in the computational basis:
\begin{equation}
\ket{\psi} = c_0\ket{00} + c_1\ket{11}+ c_2\ket{22},
\end{equation}
with $c_n\in\real$. A simple calculation leads to
\begin{equation}\label{eq:probqutrit}
\Pr(a_j\equiv b_k+\delta)= \frac{1}{9}\,\sum_{n,m=0}^3 c_n\,c_m\,
\cos\left[(n-m)\,\left(\alpha_j+\beta_k+\frac{2\pi}{3}\delta\right)\right].
\end{equation}
We can now use Equation~\eqref{eq:probqutrit} in the Inequality~\eqref{eq:i3} and optimize with respect to
$\alpha_j$ and $\beta_k$ for any possible state. For the maximally
entangled state ($c_n=\frac{1}{\sqrt{3}}$), the best choices are
$\alpha_1=0$, $\alpha_2=\frac{\pi}{3}$, $\beta_1=-\frac{\pi}{6}$ and
$\beta_2=\frac{\pi}{6}$, and one obtains $\text{\textit{CGLMP}}_Q(\ket{\Psi_3})=
4(2\sqrt{3}+3)/9\approx 2.873$ \cite{cglmp}. However, {Ac\'\i n},
Durt, Gisin and Latorre~\cite{adgl02} found that, {\em for the very
same choice of settings}, another state gives a higher violation.
Specifically, the violation $\text{\textit{CGLMP}}_Q(\ket{\psi(\gamma)})=1 + \sqrt{11/3}\approx 2.915$ is
obtained for the non-maximally entangled state
\begin{equation}\label{eq:psigamma}
\ket{\psi(\gamma)} =
\frac{1}{\sqrt{2+\gamma^2}}\,\big(\ket{00}+\gamma\ket{11}+\ket{22}\big),
\end{equation}
where $\gamma=(\sqrt{11}-\sqrt{3})/2\approx 0.792$. It should be
noted that no larger violation can be obtained by exploring all
possible states and settings and that the violation for the
maximally entangled state is also optimal. In conclusion, for a
system composed of two qutrits, the unique Bell inequality which
uses two settings per qutrit also exhibits an anomaly in its measure
of non-locality.

\subsection{Two other measures}

The picture becomes even more involved when, keeping the two
qutrits and the two settings, one explores other measures of
non-locality than the violation of the Bell inequality.

The first of these measures considered here is the classical
relative entropy, or Kullback-Leibler (K-L) distance, which measures
the ``distance'' between two probability distributions $P$ and $P'$
in terms of information. More precisely, the K-L distance is the
average amount of support in favor of $P$ against $P'$, when data
are generated using $P$. Explicitly,
\begin{equation}
{\cal{D}}(P||P') = \sum\limits_{z} P(z) \log \left(
\frac{P(z)}{P'(z)} \right)
\end{equation}
where the $z$s are the possible outcomes or events. Using this
notion, we can define another natural measure of non-locality: {\em
the K-L distance of the non-local probability distribution under
study to the closest local probability distribution}, that is
\begin{equation}
D(P_{NL}) = s\min_{P_L\in {\cal{L}}} {\cal{D}}(P_{NL}||P_L)
\end{equation}
where $P_L$ and $P_{NL}$ are respectively the local and non-local
probability distributions and ${\cal{L}}$ is the polytope of local
correlations. This measure of non-locality was studied
in~\cite{agg05}, to which we refer the reader for all details. An
interesting result stands out: when optimizing over the settings,
the maximally entangled states can give rise to non-local
correlations such that $D= 0.058$ bits, no more. Nevertheless, the
largest value, $D= 0.077$ bits, can be obtained for the correlations
generated from a non-maximally entangled state of the form
\eqref{eq:psigamma}, but here with $\gamma\approx 0.642$.

The other new measure of non-locality derives from a very recent
idea: can one demonstrate the security of key
distribution\footnote{A cryptographic paradigm where Alice and Bob
want to extend to number of secret bits they share with one another,
but with no one else.} whenever Alice and Bob share a non-local
distribution, even against an adversary who would not be limited by
quantum mechanics but only by causality? The partial answers found
to date to this question are positive~\cite{kent,agm05,sgbam06}.
Then one can define a measure of ``useful'' non-locality: the most
non-local state is the one from which one obtains the correlations
that ensure the {\em highest rate of extractable secret key against
an adversary limited only by causality}. Under several assumptions
that we cannot review here, the most non-local state of two qutrits
according to this criterion is again of the form
(\ref{eq:psigamma}), with a value $\gamma\approx 0.987$, different
from the previously encountered ones \cite{sgbam06}.

\subsection{Summary}

We have discussed three measures of non-locality, all of which show
an anomaly in the case of qutrits when the freedom of the
measurement is restricted to two settings per qutrit. Interestingly,
the most non-local states are always of the same form, but with
different numerical coefficients according to the different
measures. Actually, numerical evidence shows that this observation
is true beyond the case of qutrit, that is for any $d>3$
\cite{adgl02,agg05,sgbam06}. Thus, it seems that the scenario where
the participants share two qubits is the {\em only} case where the
anomaly does not appear for all the measures studied here.

The main restriction of the results reviewed in
Section~\ref{sec:basic},~\ref{sec:dlh} and~\ref{sec:klcrypto} is the
limitation of the number of settings to two per party. It is still
possible---although improbable in our opinion---that the anomaly
will resorb when increasing the number of settings. Needless to say,
this is a hard open problem.

\section{Simulation of entanglement with non-local
resources}\label{sec:sim}

In the previous Section, we have considered measures of non-locality
which are directly inspired by experiments or that are related to
specific configurations, fixed number of settings. A novel approach
was developed in the recent few years with the advent of computer
scientists in the field. A different measure of non-locality has
been put forth: the simulation of quantum correlations by classical
non-local resources. The intuition behind this concept is ``the more
non-local a state is, the more non-local resources will be required
to generate the same correlations''. We review this framework in the
present Section.

\subsection{Non-local resources}

Bell has showed that one cannot simulate the correlations arising
from quantum states using only local resources. For such a negative
statement, one just has to find a \emph{gedanken} experimental setup
in which a contradiction arises, and this is precisely what a Bell
inequality does\footnote{Incidentally, it is remarkable that a
contradiction can be already found using only two measurements per
particle.}. Given that local resources are not enough, how much of
\emph{non-local resources} must be added in order to simulate the
quantum correlations? The goal here is to reproduce the quantum
mechanical results for all possible measurements, not only for a
finite number of them. This question was first formulated in 1992 by
Maudlin~\cite{maudlin92}. However, the result was published in a
philosophical proceeding and went unnoticed from the quantum
information community for several years. Independently, Brassard,
Cleve and Tapp revived the field in 1999 by improving on the result
of Maudlin~\cite{tcb99}\footnote{It is important to note that
Brassard, Cleve and Tapp were blissfully ignorant of Maudlin's
paper. Else they might not even have attempted to prove their
theorem, since an argument against it was formulated in Maudlin's
paper. Moreover, Steiner also published independently a result along
these lines, merely several days after Brassard, Cleve and Tapp.
However, Maudlin already had a better result then he.}. These
authors found that one could simulate all quantum correlations that
can arise from the state $\ket{\Phi^+}$, see
Equation~\eqref{eq:singlet}, by adding eight bits of communication
to an infinite amount of LHVs. This result is already noteworthy,
because it was not evident a priori that a finite amount of
communication could do\footnote{Maudlin actually claimed that it
could not been done.}. More noteworthy still, after some
improvements, in 2003 Toner and Bacon provided an explicit model in
which \emph{a single bit of communication} added to local variables
is enough to reproduce the correlations of
$\ket{\Phi^+}$~\cite{tb03}.

Bits of communication are clearly a non-local resource, since they
can be used to signal information. Nonetheless, they also have an
unpleasant feature: they can be used to signal information. It is a
well known fact that quantum correlations cannot be used to
communicate. No matter how spooky the correlations appear, they
remain causal. It would be much more elegant to find another
non-local resource for which the no-signaling condition holds by
definition. In 2004, Cerf, Gisin, Massar and Popescu (CGMP) found
such a resource~\cite{cgmp04}: the \emph{non-local box} (NLB)---a
mathematical object invented ten years before by
Tsirelson~\cite{tsirelson} and independently by Popescu and
Rohrlich~\cite{pr94} to solve related but different problems. The
NLB is a virtual device that has two input-output ports such that if
Alice inputs a bit into her end, the NLB gives her a uniformly
random bit, likewise for Bob. The non-locality appears from the fact
that the exclusive-OR (sum modulo 2) of the outputs is always equal
to the logical AND (product) of the inputs. CGMP proved that a
single use of the NLB added to local variables allows to simulate
perfectly the correlations of $\ket{\Phi^+}$.

In summary, the statistics arising from measurements on the
maximally entangled state of two qubits can be simulated exactly by
adding to the local variables either one bit of communication or the
even weaker resource called the NLB. These results are nicely
re-derived in a unified way in~\cite{degorre}.

\subsection{The anomaly}\label{sec:resanom}

One might expect that the simulation of non-maximally entangled
states of two qubits would follow as an easy generalization of the
simulation of the maximally entangled state. Quite the opposite is
true. The state-of-the-art is as follows.

For non-local resources that allow signaling: correlations arising
from non-maximally entangled states can be reproduced using two bits
of communication~\cite{tb03}---actually a weaker resource, the
Oblivious-Transfer Box which is a well-known primitive of
information science, is sufficient~\cite{gpsww06}. It is not known
whether even weaker resources (ultimately, one bit) could do.

Using the non-signaling NLB, a sharper evidence of the anomaly has
been found: a single use of the NLB is provably {\em not sufficient}
to reproduce the statistics of non-maximally entangled states of two
qubits \cite{bgs05}.

One might conceive another physically meaningful non-local resource:
a single instance of the maximally entangled state of two qubits.
Even with this quantum resource, it has been showed that a maximally
entangled state and an infinite amount of LHVs are not sufficient to
simulate a non-maximally entangled state~\cite{g}. In the case where
we are restricted to von Neumann measurements on the singlet state,
the proof follows from the fact that we can simulate any von Neumann
measurement on a maximally entangled state with one NLB and shared
randomness, while these resources are not sufficient to simulate
measurements on a non-maximally entangled state. The general case
where we are allowed to perform any POVM on the maximally entangled
state is more intricate and it is not within the scope of this paper
to cover this proof.

In conclusion, the anomaly of non-locality shows up also for the
``amount of non-local resources''.

\section{Hardy's theorem}\label{sec:hardy}
Hardy used a different type of construction then Bell in order to
show the non-local nature of entanglement~\cite{hardy92}. Although
not a measure of non-locality, we think it is fitting to present it
here, for it also reveals an interesting property of entanglement.
We will present here Brassard's rendition~\cite{brassard} of Hardy's
proof for simplicity, in the same fashion as in~\cite{methot05}.
Lets say that Alice and Bob share the state $(\ket{01} + \ket{10} +
\ket{11})/\sqrt{3}$ along the $z$ axis. Let say that Alice and Bob
are now given the choice of performing either the $\sigma_z$ or the
$\sigma_x$ measurement. According to quantum mechanics, if Alice and
Bob are to measure $\sigma_x\otimes\sigma_x$, then they will receive
the output $(-1,-1)$ with probability $1/12$. Let us now assume that
the state has LHVs that will produce a $(-1,-1)$ output on a
$\sigma_x\otimes\sigma_x$ measurement. From the criteria of locality
and realism, we now have that any local $\sigma_x$ measurement on
this particular state will produce the output $-1$. Let us now see
what happens if Alice and Bob are to measure
$\sigma_x\otimes\sigma_z$ or $\sigma_z\otimes\sigma_x$. From the
predictions of quantum mechanics, the state should never be allowed
to produce $(-1,-1)$ as output. Once again according to the criteria
of locality and realism, we are forced to conclude that upon a local
$\sigma_z$ measurement, the state will produce $+1$ as output.
Therefore, a $\sigma_z\otimes\sigma_z$ measurement on this
particular instance is bound to output $(+1,+1)$, which is forbidden
by quantum mechanics. So in order for the LHV theory to output
$(-1,-1)$ on a $\sigma_x\otimes\sigma_x$ measurement with a non zero
probability, it will also output $(+1,+1)$ on a
$\sigma_z\otimes\sigma_z$ measurement with non zero probability, in
clear contradiction with the predictions of quantum mechanics. It is
very interesting to note that this construction works with almost
any states of two qubits. Among the exceptions, we count separable
states and maximally entangled states. Thus, we have an approach to
non-locality, which works only if the state is non-maximally
entangled. It should be noted that this fact does not depend on
Hardy's construction, but that it is also true for any Hardy-type
proof of non-locality~\cite{methot}.

\section{Perspectives on non-locality}\label{sec:conc}

\subsection{The (partial) end of the story...}

It is through quantum theory that physicists have discovered
non-locality, and the only channel known in nature that allows to
distribute non-local correlations are entangled quantum particles.
It is therefore understandable that entanglement and non-locality
have been essentially identified for many years. Results like
Eberhard's \cite{eberhard93} were either ignored, or considered as a
curiosity. When ten years later Ac\'{\i}n, Durt, Gisin and
Latorre~\cite{adgl02} found the anomaly for Bell inequalities for
qutrits, the strangeness of the result did not hit immediately,
because it still might have been an anomaly of the Bell inequality
(the fact that the CGLMP inequality is unique was proved later).

Four years later however, we cannot escape the evidence:
\emph{non-locality and entanglement are not only different concepts,
but are really quantitatively different resources}. This discovery
has been triggered by people working in quantum information, mostly
with a perspective coming from information science.

A remark is required before sketching further interesting research.
We have listed a large number of ``measures of non-locality'' such
as violation of Bell inequalities, Kullback-Leibler distances,
extractable secret key rate, robustness against the detection
loophole and simulation of entanglement with signaling and
non-signaling resources. They all seemed to be infected with the
same anomaly: maximally entangled states are not maximally
non-local. Why couldn't we simply accept the measure of entanglement
itself, see Equation~\eqref{entg}, as a measure of non-locality?
Then the anomaly would disappear by construction. The answer to this
question is subtle and instructive. The quantity
${\cal{E}}\left(\ket{\psi_{AB}}\right)$ is the fraction of maximally
entangled states of two qubits that one can extract out of a given
entangled state with the procedure called
\emph{distillation}~\cite{bbpssw96}. In distillation, one supposes
that Alice and Bob share $N$ copies of an entangled state
$\ket{\psi_{AB}}$, that they can make any local operations involving
as many of their particles as they wish and that they can
communicate through a classical channel as much as they desire. The
objective is to convert as many non-maximally entangled states
(states with low local entropy) into maximally entangled states
(states with maximal local entropy), while maybe losing a few states
in the process. This is highly similar to the notion of
\emph{block-coding} in computer science. In block-coding, one is
interested in transforming a long string of bits, which do not
necessarily have a high entropy, into a shorter string with every
bit having maximum entropy. Therefore, one can see quantum
distillation as a form of entanglement block-coding. However,
non-locality experiments are not usually performed in such a way:
Alice and Bob receive first one pair and measure it, then a second
pair and measure it, and so on. In other words, non-locality appears
in nature without the need of block-coding. Now the reader can read
back in the article and notice indeed that none of the measures of
non-locality listed above require block coding (in particular, the
probability distributions one is working with are those of repeated
single-pair measurements). This is the reason why we cannot use
${\cal{E}}\left(\ket{\psi_{AB}}\right)$ as a measure of
non-locality. Therefore, if we take the reasonable assumption that
the non-locality measures presented here actually measure
non-locality in some specific way, we can then conclude that
non-locality and entanglement are different \emph{concepts}.

To our knowledge, the first example that demonstrated that
non-locality and entanglement where different \emph{resources} came
from~\cite{bgs05}. It is well known from Tsirelson's
bound~\cite{cirelson80} that quantum mechanics cannot achieve a
perfect simulation of the NLB, embodiment of non-locality. This is
true independently of the amount of entanglement shared by the
participants. However, Cerf, Gisin, Massar and Pospescu have shown
that a single NLB is sufficient to simulate bipartite measurements
of a maximally entangled state\cite{cgmp04}. Nonetheless, as pointed
out in Section~\ref{sec:resanom}, it was proven in~\cite{bgs05},
that at least two NLBs are required to simulate some non-maximally
entangled state of two qubits. This was the first hint that NLBs are
not strictly more powerful resources than entanglement in a
one-to-one comparison. The general proof was laid down
in~\cite{bm05}, where the authors have shown quantum correlations
that require an exponential, in the number of maximally entangled
states, amount of NLBs to simulate. The added power of entanglement,
when taken in large numbers, comes from the fact that we can
entangle two $\ket{\Phi^+}$ together.

\subsection{...and the many open perspectives}

We conclude on a non-exhaustive list of open problems:
\begin{enumerate}

\item Have we established an exhaustive list of measures of
non-locality? Or, could we find others, and if yes, will they all
show the anomaly?

\item What is the physical reason of the anomaly?

\item If entanglement and non-locality are really different
resources, can one find a protocol in which non-locality (and not
entanglement) must be optimized?

\item How does the picture generalize to multipartite
entanglement, where even the notion of maximal entangled is not
uniquely defined?

\item Another direction of research would be to examine the properties of non-locality measures,
such as additivity.

\end{enumerate}

As it turns out, research in non-locality is not only the pleasure
of inventing new Bell inequalities: more than forty years after
Bell's work, there are very basic issues which are not well
understood.

\section*{Acknowledgements}

We thank Antonio Ac\'{\i}n, Hugues Blier, Anne Broadbent, Elham
Kashefi and Nicolas Gisin for stimulating discussions. We
acknowledge financial support from the European Project QAP.

\end{document}